\documentclass{article}

\usepackage{PRIMEarxiv}

\usepackage{authblk}
\usepackage{bm}

\usepackage[utf8]{inputenc} 
\usepackage[T1]{fontenc}    
\usepackage{hyperref}       
\usepackage{url}            
\usepackage{booktabs}       
\usepackage{amsfonts}       
\usepackage{nicefrac}       
\usepackage{multirow}       
\usepackage{microtype}      
\usepackage{lipsum}
\usepackage{fancyhdr}       
\usepackage{graphicx}       
\graphicspath{{media/}}     

\usepackage{enumitem}

\usepackage{caption}
\usepackage{subcaption}
\usepackage{amsmath}
\usepackage{comment}

\usepackage{hyperref}
\newcommand{\x}{\times}

\newcommand{\R}{\mathbb{R}}

\newcommand{\dsum}{\displaystyle\sum}

\usepackage{xcolor}
\usepackage{soul}

\title{Neural Fields for Highly Accelerated 2D Cine Phase Contrast MRI
}

\author[1]{Pablo Arratia}
\author[2]{Martin J. Graves}
\author[2]{Mary McLean}
\author[3]{Carolin Pirkl}
\author[4]{Carola-Bibiane Schönlieb}
\author[3]{Timo Schirmer}
\author[3]{Florian Wiesinger}
\author[1]{Matthias J. Ehrhardt}

\affil[1]{Department of Mathematical Sciences, University of Bath, Bath, UK}
\affil[2]{Department of Radiology, University of Cambridge, Cambridge, UK}
\affil[3]{GE HealthCare, Munich, Germany}
\affil[4]{Department of Applied Mathematics and Theoretical Physics, University of Cambridge, Cambridge, UK}

\begin{document}
\maketitle

\begin{abstract}

2D cine phase contrast (CPC) MRI provides quantitative information on blood velocity and flow within the human vasculature. However, data acquisition is time-consuming, motivating the reconstruction of the velocity field from undersampled measurements to reduce scan times. In this work, neural fields are proposed as a continuous spatiotemporal parametrization of complex-valued images, jointly modeling magnitude and phase across multiple echoes to enable velocity estimation, and leveraging their inductive bias for the reconstruction of the velocity data. Additionally, to compensate for the oversmoothing tendency observed in neural-field reconstructions under severe undersampling, a simple voxel-based postprocessing step is introduced. The method is validated numerically in Cartesian and radial k-space with both high and low temporal resolution data. This approach achieves accurate reconstructions at high acceleration factors, with low errors even at 32$\times$ and 64$\times$ undersampling for the high temporal resolution data, and 16$\times$ for the low temporal resolution data, and consistently outperforms classical locally low-rank regularized voxel-based methods in both flow estimates and anatomical depiction.    

\end{abstract}


\section{Introduction}

2D CPC MRI encodes not only anatomy but also velocity information in the phase of the MR signal~\cite{pelc1991phase}. This technique acquires data from a slice perpendicular to relevant vessels, such as the aorta, to recover a spatiotemporal scene. At reconstruction, phase data is converted into quantitative velocity data and flow data by virtue of measuring the vessel area~\cite{wymer2020phase}. This description is clinically relevant for assessing conditions such as regurgitation, aortic stenosis, and coarctation, among others~\cite{gatehouse2005applications, stalder2008quantitative}. In 2D CPC, k-space data from two temporally adjacent gradient echoes with different velocity encodings, $f^0$ and $f^1$, are acquired. We seek to recover the two complex-valued images $u^0,u^1$ that explain the measured data. Their magnitudes share the same anatomy, whilst their phase difference is directly proportional to the velocity. The method is also known as 2D flow MRI or velocity-encoded MRI. 

Acquiring $f^0$ and $f^1$ is done over multiple cardiac phases and respiratory cycles, leading to acquisition times of several minutes. This has motivated novel methods in the MRI community to retrieve $u^0$ and $u^1$ from undersampled data~\cite{jung2008highly, kim2012accelerated, holland2010reducing}. In parallel imaging, the data is collected from multiple receiver coils with spatially varying sensitivities~\cite{pruessmann1999sense}; in compressed sensing, redundancy in the image data is exploited by seeking sparsity in suitable domains using a regularizer in a variational model~\cite{lustig2007sparse, benning2014phase}. A popular example for dynamic MRI is the locally low-rank (LLR) regularizer, which penalizes the rank of the Casorati matrix over small patches of the scene~\cite{zhang2015accelerating}.

In the last decade, neural fields have garnered attention as a mesh-free, differentiable, biased toward smoothness, and compact representation of a scene~\cite{xie2022neural}. A neural field parametrizes a function with a deep fully-connected neural network, whose input is a point in space $\bm{x}$ and output is, for instance, the intensity of the image at that point. The sought quantity is then implicitly defined by the weights and architecture of the network, which has motivated the use of the term Implicit Neural Representation as well~\cite{sitzmann2020implicit}. Neural Radiance Fields (NeRF)~\cite{mildenhall2021nerf} constitute a popular example, where a novel-view synthesis problem is solved with a neural field that maps an input location and a view angle into a vector specifying the RGB color and opacity of the scene. These have also been used for medical imaging tasks, such as computed tomography~\cite{zang2021intratomo} and MRI~\cite{xu2023nesvor}. We refer to~\cite{molaei2023implicit} for an extensive survey on neural fields for medical imaging.  The previous works have been extended to dynamic settings by including the time as an additional variable to the network's input~\cite{pumarola2021d, lozenski2022memory}. In the context of dynamic MRI, most works map a spatiotemporal point $(\bm{x}, t)\in\Omega\times[0,T]$ to real and imaginary parts of the complex-valued image~\cite{feng2025spatiotemporal, huang2025subspace, catalan2025unsupervised, spieker2025pisco}. In this context, neural fields have been successful in incorporating time regularity due to their inductive bias that promotes smoothness in time~\cite{rahaman2019spectral}.


In this work, we propose using an implicit representation to parametrize both images $u^0, u^1$. We validate our method on two datasets with different temporal resolutions, with Cartesian and radial sampling, and at several acceleration factors. In particular, we go as high as 32$\x$ and 64$\x$ acceleration factors. Additionally, we compare our method against classical LLR voxel-based regularized methods. We now summarize our main contributions:
\begin{itemize}
    \item CPC-specific neural field formulation. We propose a continuous spatiotemporal neural field that jointly models magnitude and phases across two velocity-encoded echoes, enabling direct quantitative flow estimation. In contrast to prior dynamic neural field reconstructions, phase accuracy and echo coupling are explicitly enforced and are central to the reconstruction objective.
    \item Joint variational reconstruction of both echoes: a single optimization problem is solved to reconstruct the flow. This allows the echoes to share anatomical information through a common magnitude while preserving their relative phase for velocity estimation.
    \item Hybrid voxel–neural reconstruction: motivated by the observed oversmoothing behavior of neural fields, we introduce a hybrid formulation that combines the globally consistent spatiotemporal structure of neural fields with voxel-based refinement. This approach compensates for oversmoothing while preserving temporal coherence and phase consistency.
    \item Quantitative evaluation at high acceleration: we validate the proposed framework using flow-based metrics relevant to CPC MRI on Cartesian and radial data, demonstrating robust performance at acceleration factors up to 64× for high temporal resolution data and 16× for low temporal resolution data.
\end{itemize}

The magnitude-phase parametrization is inspired by previous works~\cite{zhao2012separate, corona2021joint, santelli2016accelerating}. Additionally, we mention that implicit representations have been used for postprocessing of reconstructed velocity-encoded data to obtain denoised and super-resolved data~\cite{saitta2024implicit, garayphysics, fathi2020super}. 


\section{Methods}

\subsection{Neural fields for 2D CPC MRI}

We represent two complex-valued time-dependent images, one for each echo, sharing the same magnitude but differing in their phases. For this, we employ a neural field that maps a spatiotemporal point $(\bm{x},t)\in\Omega_T:=\Omega\times[0,T]$ to a three-dimensional vector containing the magnitude $r$ and the two phases $\varphi^0,\varphi^1$: 
\[\begin{array}{cccl}
\Phi_{\theta}:&\Omega_T &\to&\R_{>0}\times\R^2\\
&(\bm{x},t)&\to&\Phi(\bm{x},t)=(r(\bm{x},t),\varphi^0(\bm{x},t), \varphi^1(\bm{x},t))^T.
\end{array}\]
The neural field's architecture is a simple multilayer perceptron with a Fourier feature embedding~\cite{tancik2020fourier}. We refer to Section \ref{sec:neural field architecture} for more details. In particular, we ensure the magnitude $r$ is positive by applying an exponential activation function in the corresponding neuron of the output layer.

The neural field does not have a closed form for its Fourier transform. A common approach then is to obtain a discretized image by evaluating the neural field at grid points, and then apply the discrete Fourier transform on the rasterized image. For this, we assume the domain $\Omega=[-1,1]^2$ and the time length $T=1$. This domain is then discretized with $N=N_xN_y$ points in space and $N_T$ points in time using an equispaced grid $\{\bm{x}_i\}_{i=1}^{N}\times\{t_j\}_{j=1}^{N_T}\subset[-1,1]^2\times[0,1]$. We then let $R_{\theta}, \Psi_{\theta}^0,$ and $\Psi_{\theta}^1$ to be the rasterized magnitude and complex exponential of phases:
\[\begin{array}{rl}
R_{\theta}:=&\{r_{\theta}(\bm{x}_i,t_j)\}_{i=1,\ldots,N;j=1,\ldots,N_T}\in\R_{>0}^{N\times N_T},\\
\Psi_{\theta}^0 :=& \{\exp(i \varphi_{\theta}^0(\bm{x}_i,t_j))\}_{i=1,\ldots,N;j=1,\ldots,N_T}\in\R^{N\times N_T}, \\ \Psi_{\theta}^1 :=& \{\exp(i\varphi_{\theta}^1(\bm{x}_i,t_j))\}_{i=1,\ldots,N;j=1,\ldots,N_T}\in\R^{N\times N_T}.
\end{array}\]
The two images are obtained by multiplying the magnitude and complex exponential matrices with the Hadamard product $\odot$:
\[u^0_{\theta}:=R_{\theta}\odot \Psi_{\theta}^0, \quad u^1_{\theta}:=R_{\theta}\odot \Psi_{\theta}^1.\]

Since both images share the same magnitude, we simultaneously solve for both echoes by solving one variational problem, thus, sharing the information between echoes:
\begin{equation}\label{eq:variational problem neural field}
    \min_{\theta} \mathcal{D}(\bm{K}^0u^0_{\theta}, f^0)+\mathcal{D}(\bm{K}^1u^1_{\theta}, f^1).
\end{equation}
Here, $\mathcal{D}$ is a data fidelity term that measures the discrepancy between predicted and acquired measurements, while $\bm{K}^0$ and $\bm{K}^1$ represent the imaging process, including the sensitivity maps, the Fourier transform, and the sampling scheme. In particular, $\bm{K}^0$ and $\bm{K}^1$ differ in the sampled frequencies, which are assumed to be different per echo, as explained in Section \ref{sec:experimental settings}. We refer to Section \ref{sec:variational problem} for further details regarding the variational problem. 

The evaluation of the loss in Equation \eqref{eq:variational problem neural field} requires $N\times N_T$ forward passes of the neural field. This is time-consuming and slows down optimization. Therefore, we proceed by randomly sampling $N_B=1$ frame per iteration and minimizing its distance to the data, where $N_B$ denotes the batch size. This introduces a significant speed-up for the neural field in capturing sharp edges in the image, but introduces variability throughout iterations. Consequently, larger batch sizes $1\leq N_B\leq N_T$ are used later on during training to stabilize the neural field's output.

\subsubsection{Hybrid model: a voxel-based postprocessing of neural fields}
 
Neural fields' smoothing is beneficial to gain time coherence of the scene. However, in contrast to voxel-based representations, these can struggle to capture fine details. Additionally, the chosen architecture, optimization process and the non-convex landscape of the neural field's loss do not ensure capturing all the details in the final images. This is briefly illustrated in Section \ref{sec:embedding}, where neural field with several architectures do not directly fit the desired image. We therefore propose a postprocessing step, where a voxelated solution is obtained by solving a variational problem regularized towards the neural field solution to incorporate time regularity. The goal is to obtain the best from both worlds: sharp edges from the discrete solution and time regularity from the neural field. The problem is formulated independently for both echoes as follows:
\begin{equation}\label{eq:variational problem hybrid}
   u_{\text{Hyb}}^j=\arg\min_{ u\in\mathbb{C}^{N\times N_T}} \mathcal{D}(\bm{K}^ju,f^j)+ \dfrac{\lambda_{\text{Hyb}}}{2} \|u-u_{\theta^*}^j\|_2^2, \quad j=0,1, 
\end{equation}
with $\theta^*$ denoting the weights obtained from the optimization of \eqref{eq:variational problem neural field}, and $\lambda_{\text{Hyb}}\geq 0$ is a regularization parameter weighting the influence of the neural field. This formulation can be interpreted as a Tikhonov-type regularization around the neural field solution, with $\lambda_{\text{Hyb}}$ controlling a continuous interpolation between the neural field estimate and the voxel-based SWS reconstruction. The loss is convex and smooth in $u$ and can be solved with conjugate gradient iterations. See Section \ref{sec:hybrid} for details.

\subsection{Baseline methods}

We benchmark our approach against two voxel-based methods: the Sensitivity Weighted Solution (SWS), and a locally low-rank (LLR) regularized solution. Both methods solve two independent variational problems, one per echo. The magnitude is then obtained by averaging the magnitude of both solutions, while the predicted velocity data is simply the difference of the phases. The SWS solution only fits the data term without regularization. Hence, it is expected to perform poorly for large acceleration factors, see Section \ref{sec:sws}. The LLR solution employs a locally low-rank regularizer that penalizes the rank of the Casorati matrix on small patches to enforce temporal regularity. The regularization parameter weighting this regularizer in the variational problem is denoted by $\lambda_{\text{LLR}}$, see Section \ref{sec:llr}.


\subsection{Experimental settings}\label{sec:experimental settings}

We now proceed to describe the datasets used and the retrospective undersampling for the three experiments we use to validate our method.

\subsubsection{Experiment 1. High temporal resolution dataset}

\paragraph*{Data}. The first dataset consists of fully-sampled k-space Cartesian data spanning one cardiac cycle. This data was acquired on a clinical 3T Premier MRI system (GE HealthCare, Chicago, IL) with 142$\times$142 spatial image matrix, 83 temporal frames (high temporal resolution of $\sim$12 ms), and 35 activated receive coil elements. The sequence parameters are as follows: repetition time (TR) of 5.5 ms, echo time (TE) of 3.1 ms, in-plane spatial resolution of $1.5\x 1.5$  mm$^2$, and slice thickness of 7 mm.

\begin{figure}[t!]
\centering
\subfloat[]{\includegraphics[width=0.25\textwidth]{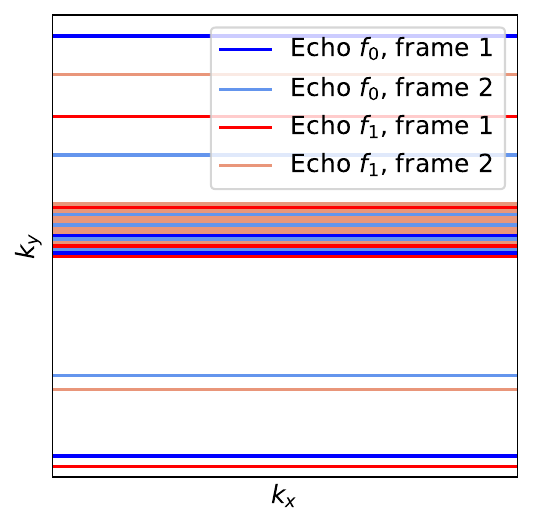}}
\subfloat[]{\includegraphics[width=0.25\textwidth]{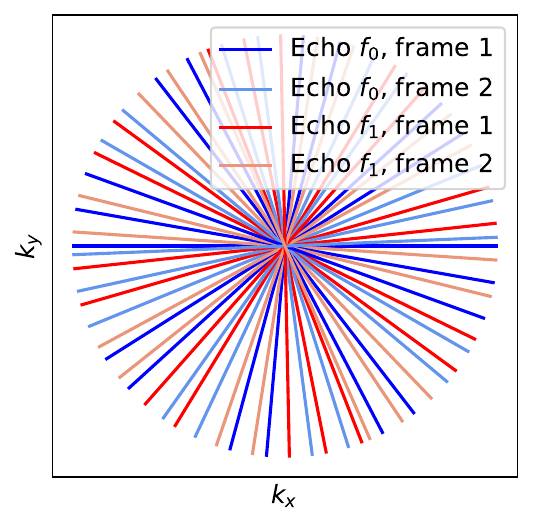}}
\caption{Retrospective variable-density and radial undersampling at factor 16$\x$. K-space lines are shown for two echoes at two different frames. The schemes ensure that different frequencies are sampled per echo at the same frame.}
\label{fig:sampling}
\end{figure}

\paragraph*{Sampling}. The fully-sampled data is retrospectively downsampled at acceleration factors of 2×, 4×, 8×, 16×, 32×, and 64×, corresponding to 71, 36, 18, 9, 5, and 3 k-space lines per frame, respectively. We employ a variable-density random sampling scheme that oversamples the 16 central k-space lines and progressively covers the remaining lines across frames. When more than 16 lines are sampled, the central region is fully covered and the additional lines are drawn from the periphery; when fewer than 16 lines are sampled, one line above and one below the center are included, with the rest drawn from the central region. To further increase measurement incoherence, different frequency lines are sampled across echoes. Moreover, we adopt a line-by-line sampling strategy in which each frame acquires a small subset of lines selected uniformly at random from those not yet sampled. Once all lines have been acquired over the course of several frames, the process restarts with the full set of lines. See Figure \ref{fig:sampling}.


\subsubsection{Experiment 2. Low temporal resolution CMRxRecon 2024 dataset}

\paragraph*{Data}. We also use data from the CMRxRecon 2024 Challenge\footnote{\url{https://cmrxrecon.github.io/2024/Home.html}}~\cite{wang2024cmrxrecon, wang2023cmrxrecon}. In particular, we use the data from 5 patients (P001--P005), in the test 2D CPC data. While the challenge provides training data for several cardiac MRI modalities (e.g., cine, tagging, and T1/T2 mapping), the 2D CPC data are not included in the training set and are treated as an unseen modality within the challenge. As a result, deep learning models developed for the challenge are not trained on CPC data. Fully-sampled Cartesian k-space data spanning one cardiac cycle is acquired on a 3T scanner (MAGNETOM Vida, Siemens Healthineers, Germany) with 144$\times$384 spatial image matrix, 12 temporal frames (low temporal resolution of $\sim$80 ms), and 10 activated receive coil elements. The sequence parameters are as follows: repetition time (TR) of 3.6 ms, echo time (TE) of 1.6 ms, in-plane spatial resolution of 1.5 $\x$ 1.5 mm$^2$, and slice thickness of 5 mm.

\paragraph*{Sampling}. The same sampling scheme used in Experiment 1 is employed up to an acceleration factor of 32$\x$. This is due to the low temporal resolution of this dataset. We highlight that the time resolution for this data (12 frames) is much smaller than the time resolution of the data used in the previous section (83 frames). Thus, worse results for the same acceleration factors are expected due to less available data.

\subsubsection{Experiment 3. Radial data}

\paragraph*{Data}. The method is further validated on radially sampled data. To achieve this, the original Cartesian data of both high and low temporal resolution datasets are interpolated using the Kaiser-Bessel kernel, as implemented in the package \emph{TorchKbNufft}~\cite{muckley:20:tah}. Experiment 3.a examines the high-temporal-resolution dataset with radial data, while Experiment 3.b investigates the low-temporal-resolution dataset with radial data.

\paragraph*{Sampling}. For the radial acquisitions, we use a golden-angle sampling strategy in which the angular step is applied across echoes. Specifically, if one echo acquires a spoke in a given direction, the next echo acquires a spoke rotated by the golden-angle increment. This rotation continues, alternating between echoes while gradually filling k-space in a highly uniform yet incoherent manner. See Figure \ref{fig:sampling}. Due to the good performance of methods on radially sampled data for large acceleration factors, we do not consider low acceleration factors. In particular, we use factors $16\x, 32\x$, and $64\x$ for the high temporal resolution data, and factors $8\x, 16\x$, and $32\x$ for the low temporal resolution data.

\subsection{Assessment}

To assess the results, we first compute a reference image as the SWS solution from the fully-sampled data. The magnitude of the solution is used to manually segment the aorta. This region is then used to compute the flow through the aorta. Finally, we report the 2-norm, $\infty$-norm, and overall flow percentage relative errors computed as described in Section \ref{sec:metrics}. We highlight that velocity encoding (VENC) information is not available for the datasets, therefore, these metrics are used as surrogates for clinically relevant quantities such as peak flow, net flow, and stroke volume.

\begin{figure*}[htbp]
\centering
\includegraphics[width=0.9\textwidth]{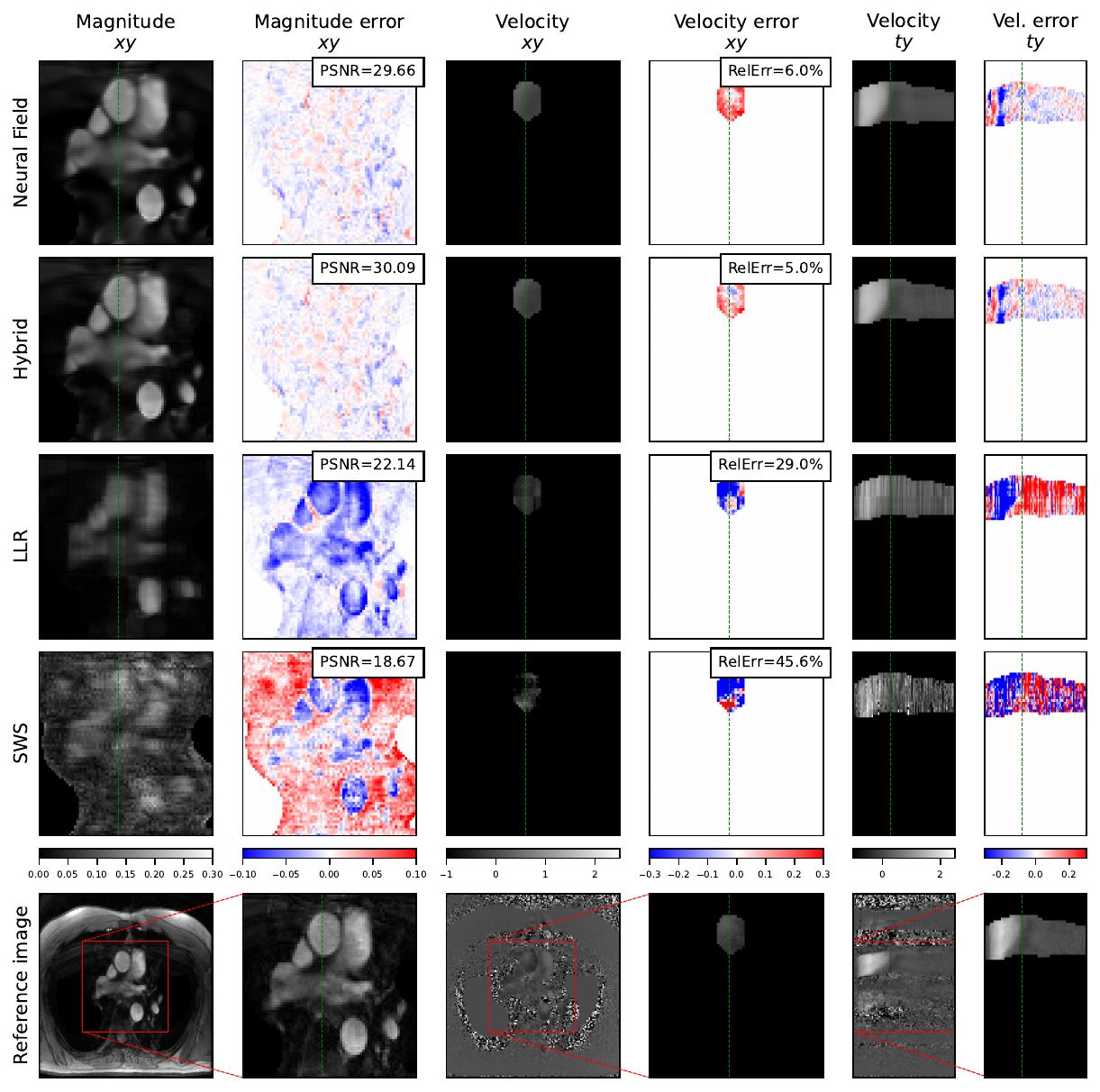}
\caption{Reconstruction results on Experiment 1 at an acceleration factor of 32$\x$. Images are zoomed in on the region of interest. Frame 30 is displayed for the $xy$ view. This is the frame where the neural field cannot capture the negative peak in the mean velocity. PSNR for the zoomed-in spatiotemporal scene and 2-norm relative error of the flow are also shown. Velocity maps are masked to the aorta region.}
\label{fig:high res 32 cart}
\end{figure*}

\section{Results}

We now present the main results. The regularization parameters for the hybrid and LLR models are obtained by performing a grid search with $\lambda_{\text{Hyb}}\in\{10^{-3},5\times10^{-3},\ldots,1 ,5, 10\}$, and $\lambda_{\text{LLR}}\in\{10^{-4}, 5\times 10^{-4}, \ldots, 5\times 10^{-2}, 10^{-1}\}$. For the LLR and hybrid methods, we use the same regularization parameters for all the acceleration factors. The chosen parameter is the one that presents the least geometric mean of 2-norm relative error on the flow across all acceleration factors. 

At initialization, the weights of the neural fields are defined using the Xavier initialization, while the biases are set to 0. The network is then optimized using the Adam optimizer with a fixed learning rate of $10^{-3}$. Computing the loss for all frames at each iteration significantly slows down optimization. Instead, at each iteration, we randomly sample $1\leq N_B\leq N_T$ frames and minimize the loss at that time. We observe that setting $N_B=1$ allows the neural field to capture edges in less time, but comes at the cost of high variability in the prediction. Therefore, we start with a batch of size $N_B=1$ and then increase it to ensure stability during optimization. For Experiments 1 and 3.a (high temporal resolution), we train for 1000 epochs with a batch size of $N_B=1$, then, then 200 epochs with a batch size of $N_B=21$, and finally, 200 additional epochs with a batch size of $N_B=42$. In particular, during optimization, the neural field never computes the entire spatiotemporal scene because we have $N_T=83$ frames. For Experiments 2 and 3.b (low temporal resolution), the neural field is trained for 5000 epochs with a batch size of $N_B = 1$, and then for 1000 additional epochs with a batch size of $N_B=12$.

Neural field experiments were run on an NVIDIA T4 GPU (16 GB), while voxel-based methods were run on an Intel Xeon CPU @ 2.20 GHz (2 cores).

\subsection{Experiment 1}\label{sec:high res Cartesian}

\begin{figure*}[htbp]
\centering
\subfloat[]{\includegraphics[width=0.33\textwidth]{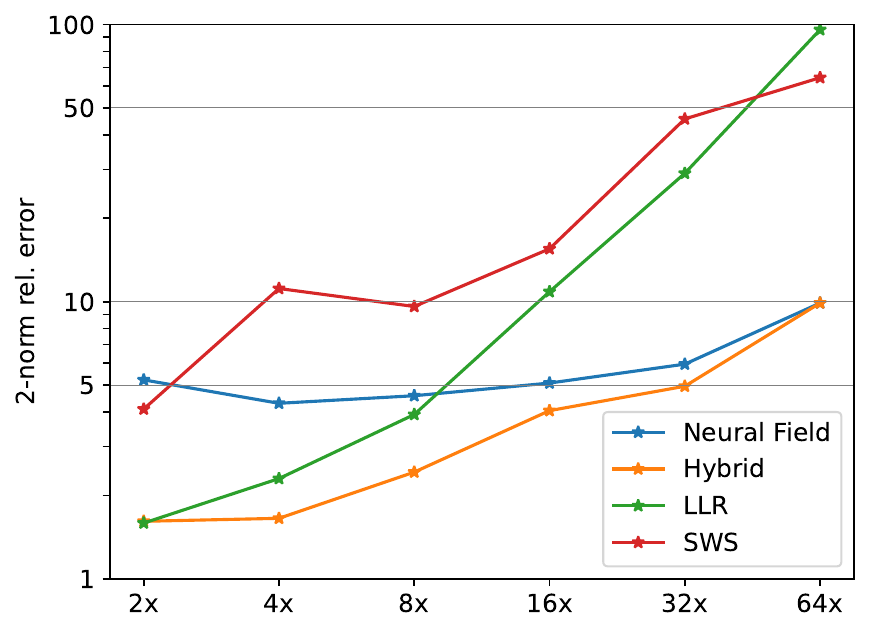}}
\subfloat[]{\includegraphics[width=0.33\textwidth]{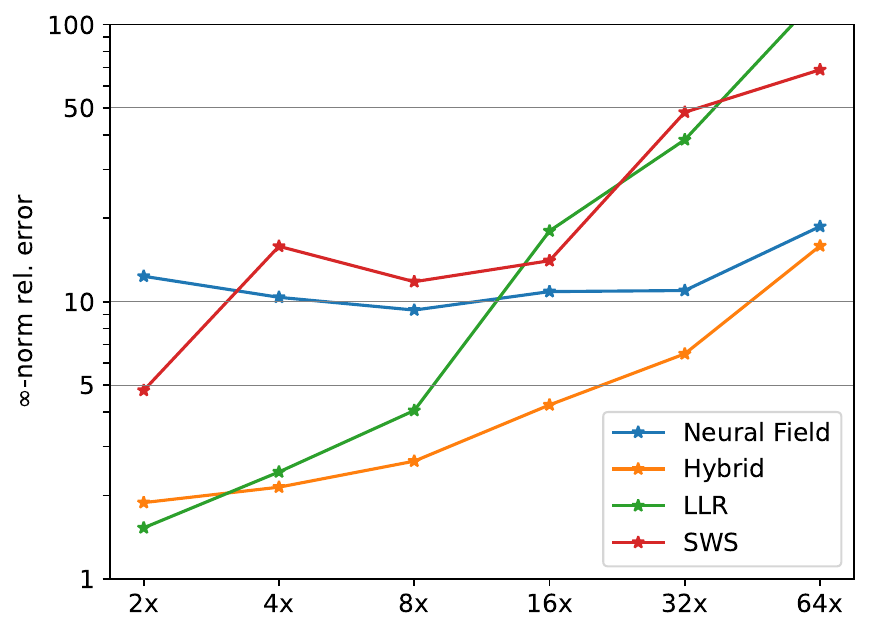}}
\subfloat[]{\includegraphics[width=0.33\textwidth]{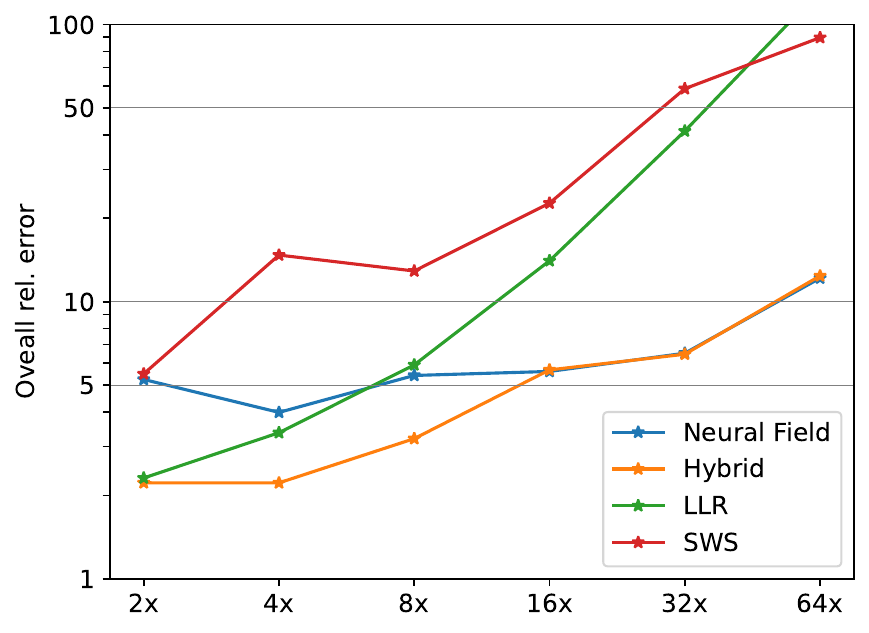}}
\caption{Flow relative errors Experiment 1 (Section \ref{sec:high res Cartesian}). Left: 2-norm relative error, center: $\infty$-norm relative error, right: overall relative error. Note the stable performance of the neural field approach across acceleration factors, reflecting its inductive bias toward smooth flow fields.}
\label{fig:high res relative error}
\end{figure*}

\begin{figure*}[t!]
\centering
\subfloat[Neural field]{\includegraphics[width=0.33\textwidth]{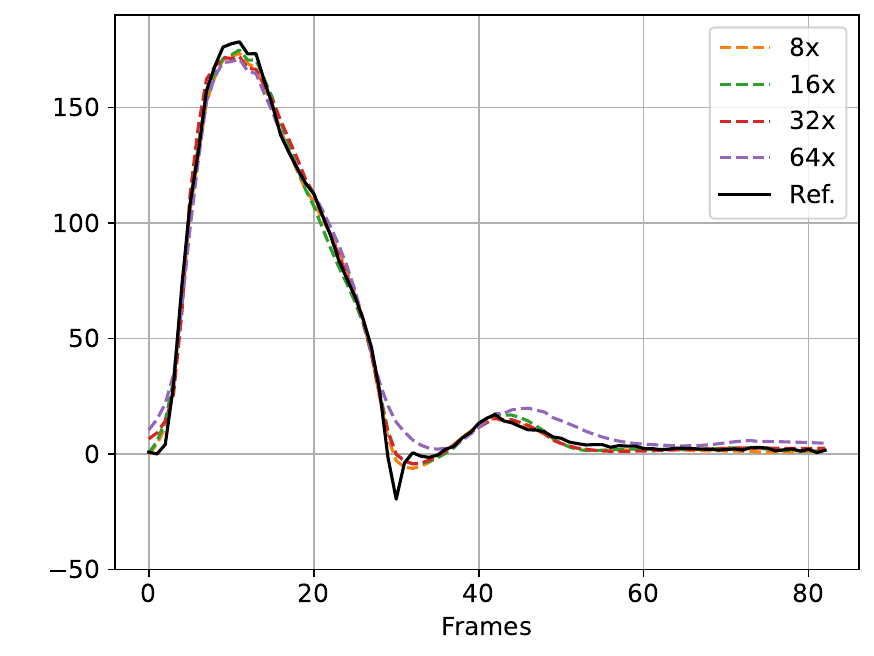}}
\subfloat[Hybrid, $\lambda_{\text{Hyb}}=10^{-1}$]{\includegraphics[width=0.33\textwidth]{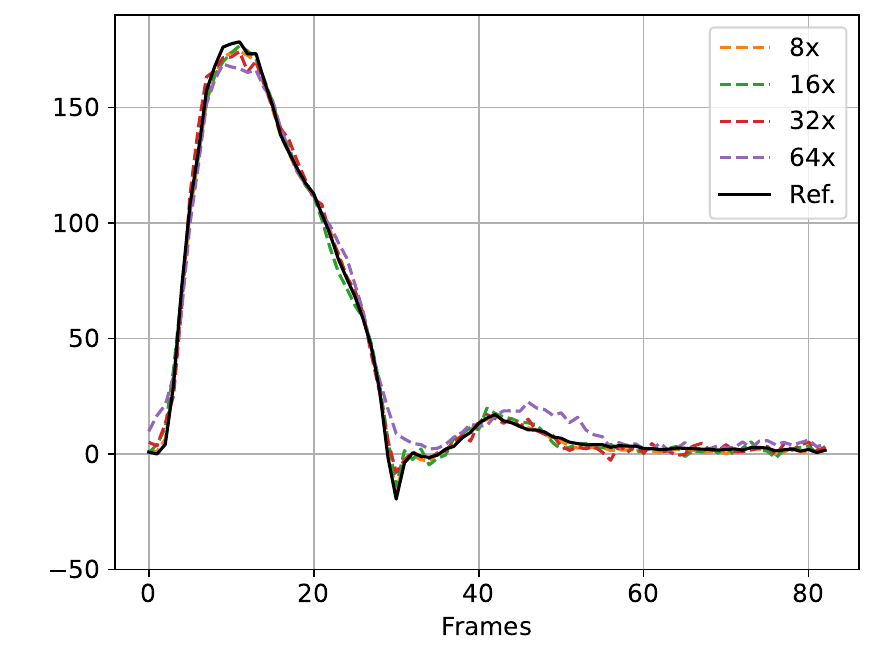}}
\subfloat[LLR, $\lambda_{\text{LLR}}=10^{-2}$]{\includegraphics[width=0.33\textwidth]{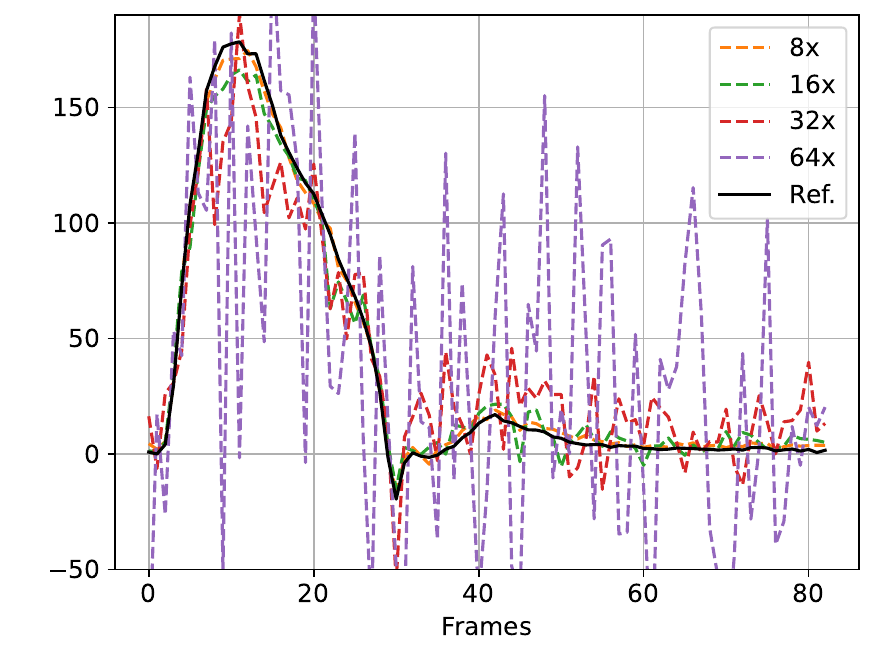}}\hfill
\subfloat[Flow errors at factor 8$\x$]{\includegraphics[width=0.33\textwidth]{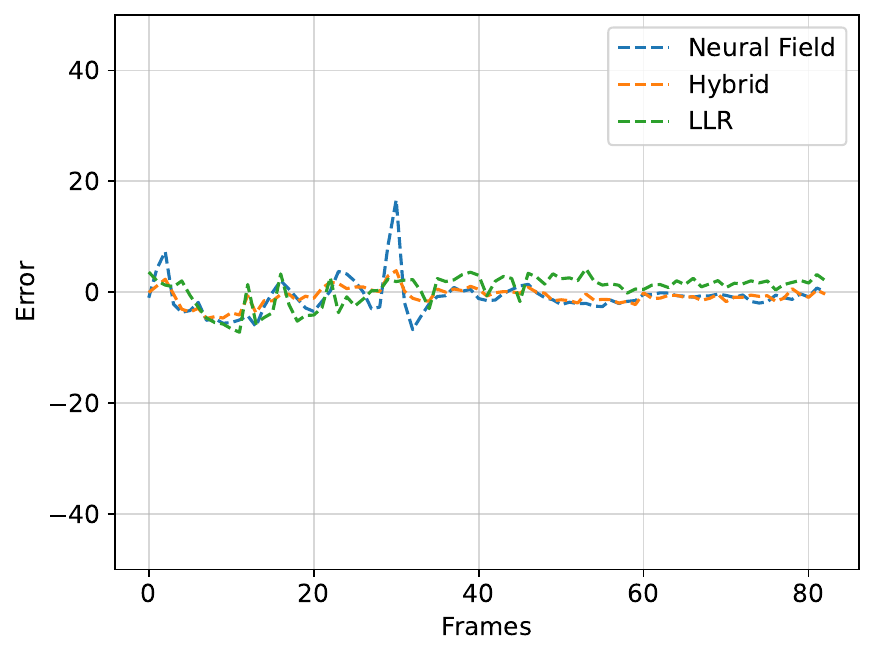}}
\subfloat[Flow errors at factor 16$\x$]{\includegraphics[width=0.33\textwidth]{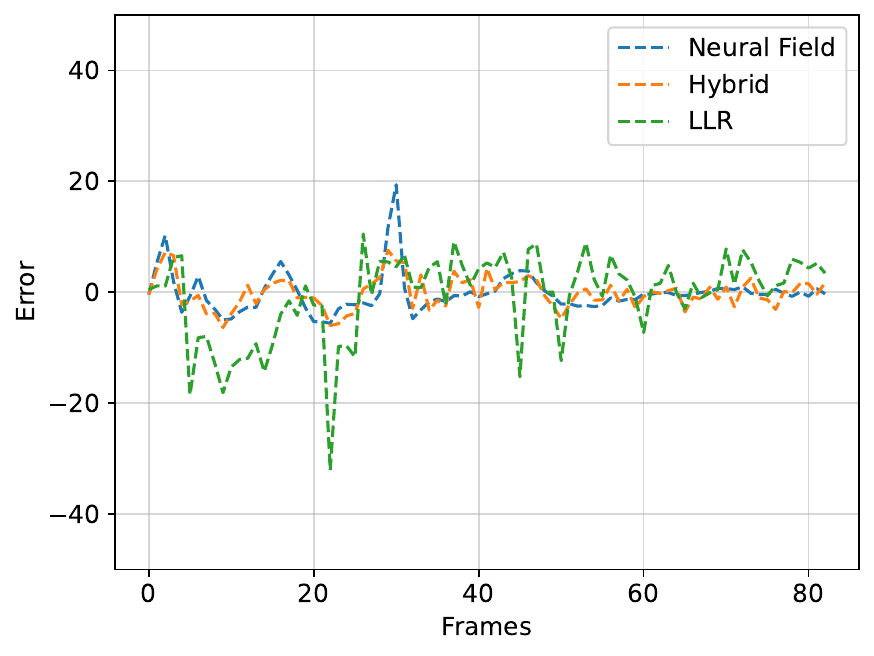}}
\subfloat[Flow errors at factor $32\x$]{\includegraphics[width=0.33\textwidth]{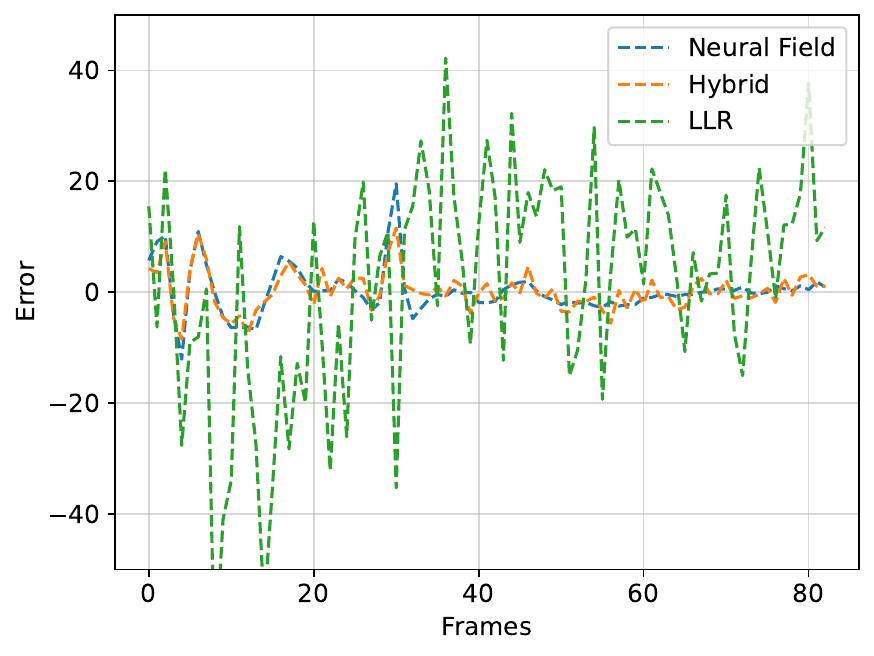}}\hfill
\caption{Top: reference flow (black) against predicted flow for neural field, hybrid, and LLR methods at different acceleration factors. The neural field struggles to capture the negative peak at frame 30, while the hybrid method does capture it except for factor $64\x$. Bottom: time-wise error of flow for neural field, hybrid, and LLR methods. The neural field presents its largest error at the negative peak in frame 30.}
\label{fig:high res flow}
\end{figure*}

We found $\lambda_{\text{LLR}}=10^{-2}$ and $\lambda_{\text{Hyb}} = 10^{-1}$ to give the best overall performance across all acceleration factors for the LLR and hybrid models, respectively. Ablation studies on these parameters are presented in Sections \ref{sec:ablation hyb} and \ref{sec:ablation llr}. Figure \ref{fig:high res 32 cart} shows the reconstruction for the neural field, hybrid, LLR, and SWS methods for Experiment 1 at an acceleration factor of 32$\x$. Despite having only $3.125\%$ of the data, the neural field and hybrid solution can capture well the region of the aorta, achieving a PSNR of 30dB approximately. The voxel-based solutions, on the other hand, introduce more artifacts and blurriness in the reconstruction. More importantly, the neural field and the hybrid method achieve a low 2-norm relative error of 6.0\% and 5.0\%, respectively. Figure \ref{fig:high res relative error} summarizes the performance of the four methods in terms of their relative errors in the flow. As expected, the unregularized SWS solution performs poorly, presenting errors above 10\% from an acceleration factor of 4$\x$. The LLR solution performance drastically drops for factors higher than 16$\x$. The neural field shows stability across acceleration factors and demonstrates clear advantages over the voxel-based ones from an acceleration of 16$\x$. For instance, it achieves a 2-norm relative error of 10\% even for a factor of 64$\x$. The comparatively stable error of the neural-field reconstructions across acceleration factors reflects the strong inductive bias of implicit representations toward smooth velocity fields, allowing the dominant flow structure to be recovered from a limited number of k-space samples. It is also observed that the neural field's 2-norm error does not go below 4\% even for low factors. This has to do with the expressive power of the neural field and its smoothness given by the network's architecture: even directly fitting the neural field to the reference image leads to a similar error in the flow. We refer the reader to Section \ref{sec:embedding} for more details. To better understand this, the predicted flows are shown in Figure \ref{fig:high res flow}. There, it is clear that the neural field struggles to capture the sharp feature occurring in frame 30. This also explains why the $\infty$-norm relative error remains large for the neural field. The situation improves when postprocessing the neural field solution with the hybrid model: the voxelated nature of this solution captures well the negative peak in frame 30 while maintaining the smoothness in the remaining frames, thanks to the regularizing effect of the neural field. This way, the hybrid model captures the best of neural fields and voxel-based representations: it retains the time coherence given by neural fields and captures abrupt changes.

\subsection{Experiment 2}\label{sec:experiment 2}

\begin{figure*}[t]
\centering
\includegraphics[width=\textwidth]{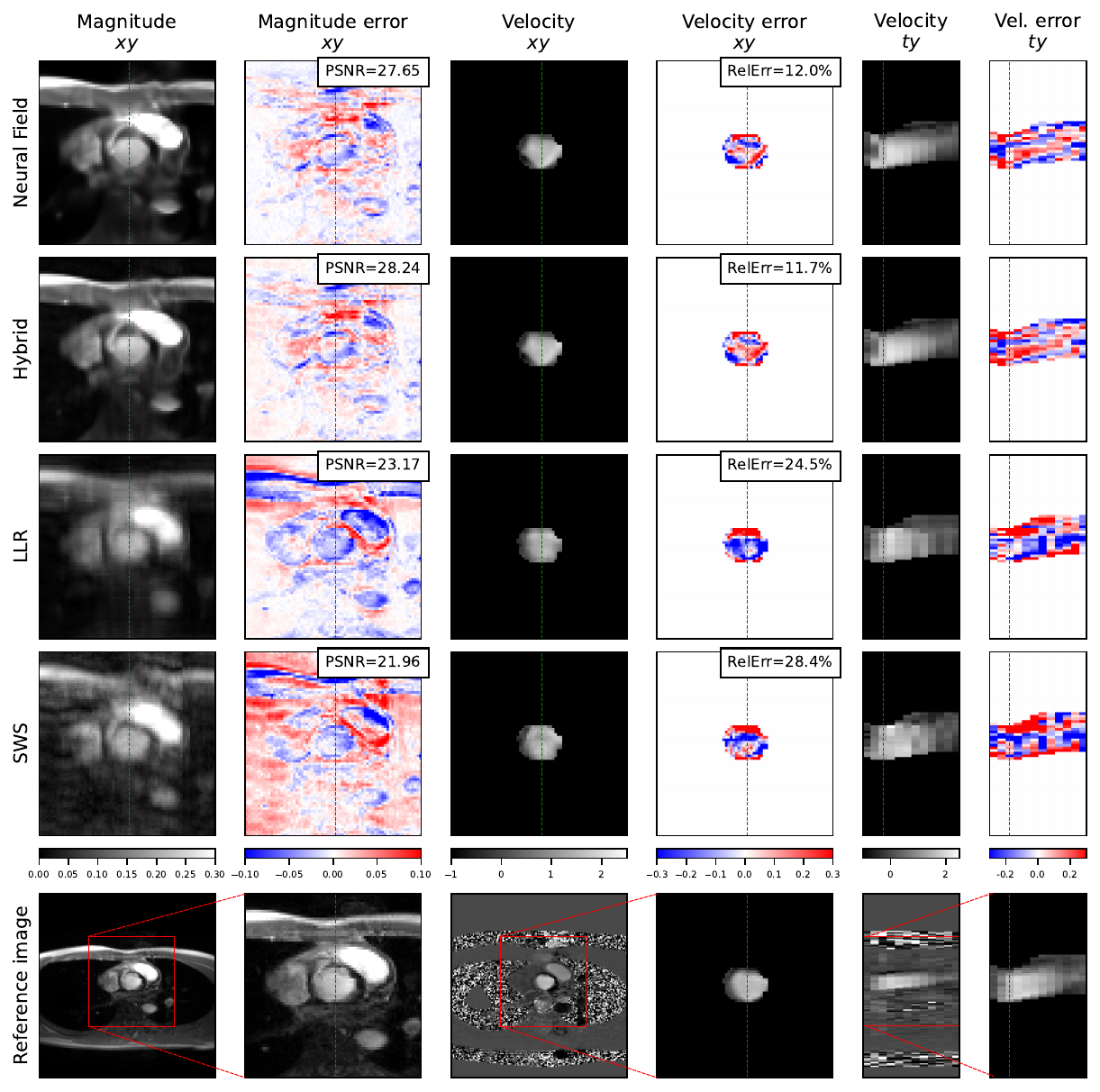}
\caption{Reconstruction results on Experiment 2 for patient P001 at an acceleration factor 16$\x$. PSNR for the zoomed-in spatiotemporal scene and 2-norm relative error of the flow are also shown. Velocity maps are masked to the aorta region.}
\label{fig:low res P001}
\end{figure*}

\begin{figure}[t]
\centering
\includegraphics[width=0.45\textwidth]{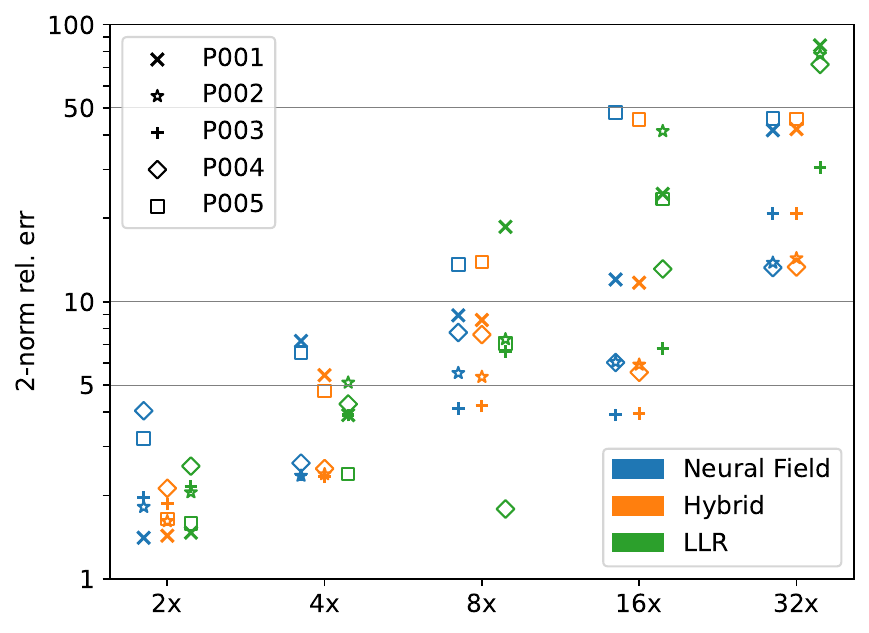}
\caption{2-norm relative errors for Experiment 2 for each method and patient. LLR reconstruction for P005 at factor $32\x$ presents an error larger than 100\%. The SWS is omitted to simplify visualization.}
\label{fig:experiment 2 2 norm}
\end{figure}

For this experiment, a grid search for the regularization parameters is performed for each patient, resulting in $\lambda_{\text{LLR}}=10^{-2}$ and $\lambda_{\text{Hyb}}=10^{-2}$ for P001; $\lambda_{\text{LLR}}=5\x 10^{-3}$ and $\lambda_{\text{Hyb}}=10^{-1}$ for P002; $\lambda_{\text{LLR}}=10^{-3}$ and $\lambda_{\text{Hyb}}=5\times10^{-1}$ for P003; $\lambda_{\text{LLR}}=10^{-3}$ and $\lambda_{\text{Hyb}}=5\x10^{-2}$ for P004; and  $\lambda_{\text{LLR}}=5\times 10^{-3}$ and $\lambda_{\text{Hyb}}=10^{-1}$ for P005. Figure \ref{fig:low res P001} shows the reconstruction using the four methods for patient P001, at an acceleration factor of 16$\x$. Similar to Experiment 1, the neural field outperforms the voxel-based baseline methods, with the hybrid postprocessing improving both the 2-norm relative error in the flow and the PSNR in the magnitude. We also observe that the low temporal resolution of this data negatively affects the neural field's performance: the relative error in Experiment 1 for factor 32$\x$ remains below 6\% (see Figure \ref{fig:high res 32 cart}), while the relative errors for neural field and hybrid methods are above 11\% for a factor of 16$\x$. The 2-norm relative errors for the five patients and neural field, hybrid, and LLR methods are displayed in Figure \ref{fig:experiment 2 2 norm}. Overall, at factors $2\x$ and $4\x$, both hybrid and LLR reconstructions present similar errors, most of them below 5\%. At factor $8\x$ the hybrid model outperforms the LLR solution for patients P001, P002, and P003. We observe that at a factor of $16\x$, the neural field and hybrid models outperform the LLR reconstruction for all patients but P005, for which large errors are observed. For the highest acceleration factor, 32$\x$, the proposed methods still perform better than the LLR solution, however, the errors in this case are too large, indicating non-realistic velocities. Finally, we mention that the hybrid model barely improves the neural field solution, meaning that the neural field can represent the reference image with high fidelity (as opposed to the situation in Experiment 1).


\subsection{Experiment 3}\label{sec:experiment 3}

\begin{figure*}[t!]
\centering
\includegraphics[width=0.9\textwidth]{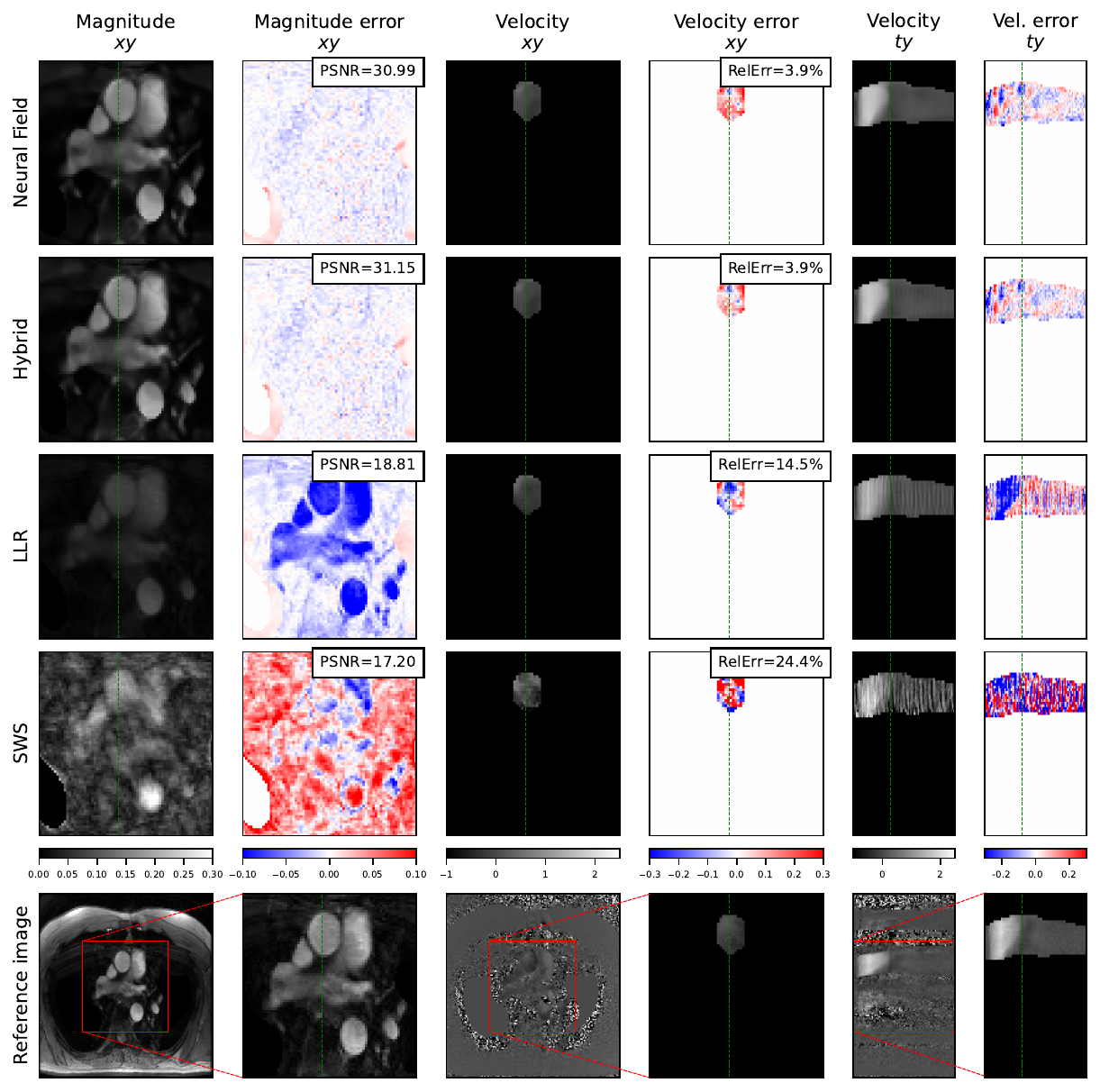}
\caption{Reconstruction results on Experiment 3.a (radial data) at an acceleration factor of 32$\x$. Images are zoomed in on the region of interest. Frame 30 is displayed for the $xy$ view. This is the frame where the neural field cannot capture the negative peak in the flow. PSNR for the zoomed-in spatiotemporal scene and 2-norm relative error of the flow are also shown. Velocity maps are masked to the aorta region.}
\label{fig:high res 32 radial}
\end{figure*}

\begin{figure*}[t]
\centering
\subfloat[2-norm relative error.]{\includegraphics[width=0.33\textwidth]{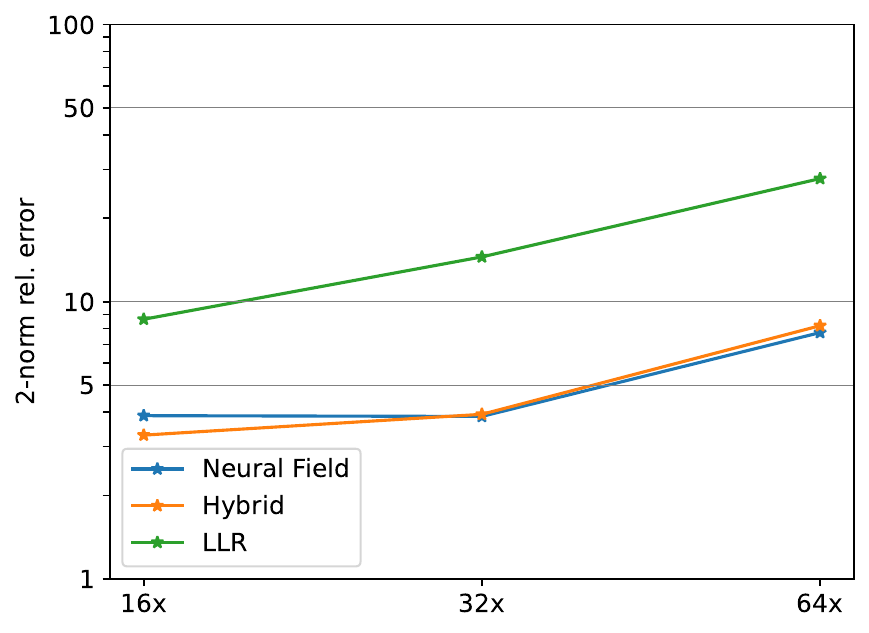}}
\subfloat[$\infty$-norm relative error.]{\includegraphics[width=0.33\textwidth]{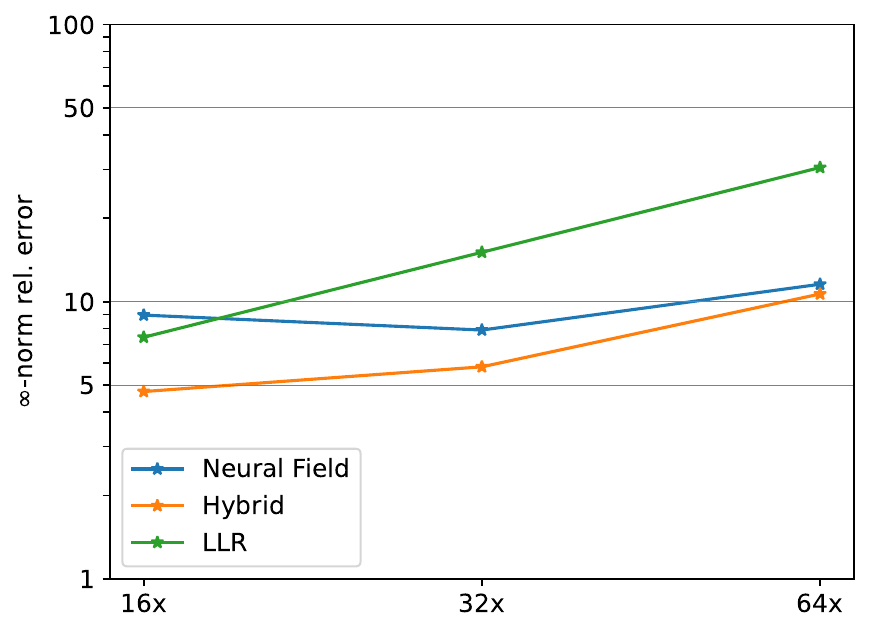}}
\subfloat[Overall flow relative error.]{\includegraphics[width=0.33\textwidth]{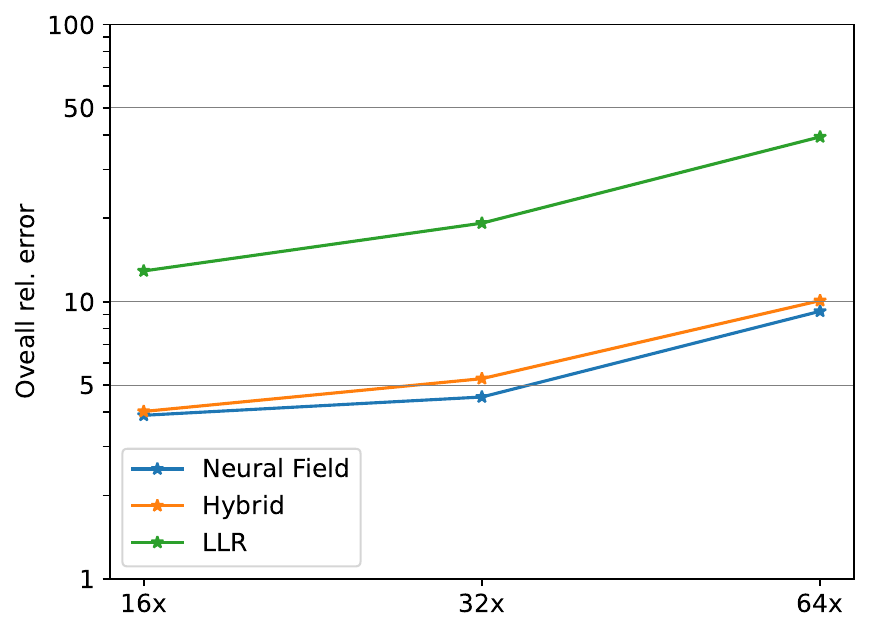}}
\caption{Flow relative errors Experiment 3.a (Section \ref{sec:experiment 3}). Left: 2-norm relative error, center: $\infty$-norm relative error, right: overall relative error.}
\label{fig:high res radial}
\end{figure*}

\paragraph*{Experiment 3.a}\label{sec:experimenta 3.a}. Motivated by the results of the previous section with neural fields achieving better results at high acceleration factors, we now replicate the experiment but for radial k-space trajectories at high acceleration factors, namely, 16$\x$, 32$\x$, and 64$\x$. Also, for simplicity, we keep the same regularization parameters obtained in Experiment 1, namely, $\lambda_{\text{LLR}}=10^{-2}$ and $\lambda_{\text{Hyb}} = 10^{-1}$. Figure \ref{fig:high res 32 radial} shows the reconstruction at an acceleration factor of 32$\x$. Compared with the Cartesian sampling counterpart in Figure \ref{fig:high res 32 cart}, we observe that the four methods improve their 2-norm relative errors in the flow. In particular, the neural field methods attain an error below 4\% with only 5 k-space lines per frame. Additionally, the LLR reconstruction shows a systematic reduction in magnitude intensity, leading to diminished contrast in the images. A similar effect is also present in Experiment 1, although to a lesser extent (see Figure \ref{fig:high res 32 cart}).  Figure \ref{fig:high res radial} shows the relative errors for all the factors. There, the hybrid model outperforms the LLR solution in all scenarios, and goes barely over 10\% error for the highest factor 64$\x$. Finally, we highlight that, as expected, radially sampled data leads to better results for dynamic MRI, achieving lower errors than the Cartesian data.

\begin{figure}[t]
\centering
\includegraphics[width=0.45\textwidth]{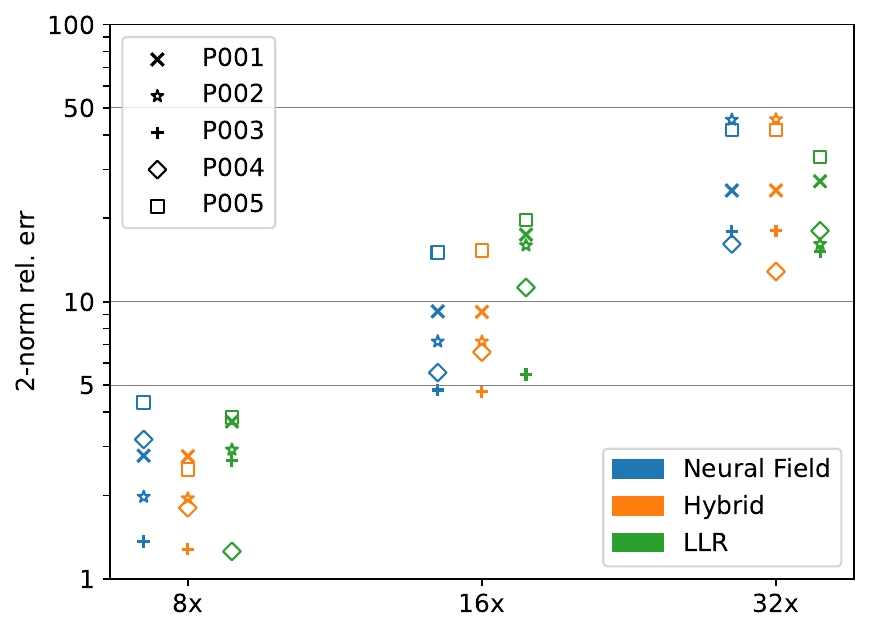}
\caption{2-norm relative errors for Experiment 3.b for each method, patient, and high acceleration factors. The SWS is omitted to simplify visualization.}
\label{fig:low res radial}
\end{figure}


\paragraph*{Experiment 3.b}. We now study the low-temporal-resolution data with radial trajectories. Again, for simplicity, we keep the same regularization parameters obtained in Experiment 2. Results are summarized in Figure \ref{fig:low res radial}. A similar trend is observed, with the neural field and hybrid methods outperforming LLR for factors $8\x$ and 16$\x$. Again, post-processing does not introduce a major improvement in the neural field's solution. We notice that for the factor 32$\x$, the LLR solution outperforms the other two methods, due to a large error in patient P002.

\section{Discussion}

The improved reconstruction accuracy indicates that the proposed neural field methods capture spatiotemporal correlations more effectively than conventional LLR methods for dynamic MRI, thus enabling high-quality blood flow estimation and image reconstruction. Our numerical experiments demonstrate that neural fields can reduce scanning times by collecting data from a few cardiac cycles, achieving errors below 4\% for radial data at an acceleration factor of 32x in Experiment 3.a. This corresponds to 3.125\% of the full data and only 5 k-space lines per frame. Numerical experiments also show larger errors in Experiment 2 than in Experiment 1, indicating that all methods benefit from measurements with fine temporal resolution. This suggests that larger acceleration factors can be used if the data is collected in small time steps.

As drawbacks for the proposed method, we note the computation times and non-convexity. Many forward passes are required when going from continuous to discrete representation since the neural field needs to be queried in $N\times N_T$ grid points. To mitigate this, we used a batch size of 1 in time (see Section \ref{sec:neural field architecture}) for most of the optimization, allowing more iterations in less time, but this routine is still slower than the voxel-based methods. For example, in Experiment 1, the neural field solution required 18 minutes to run on a GPU, whereas the LLR solution took 8 minutes per echo. The situation improves in favor of neural fields when considering radial sampling, but both methods considerably increase their time due to the use of non-uniform FFT. See Table \ref{tab:total time}. Also, Table \ref{tab:memory} shows the memory consumption and time per epoch for the neural field for different batch sizes. 
We observe a clear trade-off between memory consumption and runtime when comparing Cartesian and radial sampling strategies. For Cartesian data, the data-consistency term is evaluated by applying a full two-dimensional FFT to the predicted images followed by masking in k-space. Although the sampling mask is sparse, the computation requires forming and storing full-grid complex Fourier tensors and their associated residuals during backpropagation, leading to higher peak GPU memory usage. In contrast, for radial data we employ a Toeplitz formulation of the normal operator, with precomputed adjoint data and Toeplitz kernels, which allows the data-consistency loss to be evaluated without explicitly forming k-space predictions or residuals. This significantly reduces the number and size of intermediate tensors that must be retained for gradient computation, resulting in lower peak memory consumption during optimisation. However, the radial Toeplitz approach involves additional FFT-based convolutions on oversampled grids and more complex operator applications, which increase computational overhead per iteration. As a result, while the radial formulation is more memory-efficient, it incurs a higher runtime compared to the Cartesian FFT-based implementation. An option to accelerate neural fields is to consider hash-encodings~\cite{muller2022instant}, a novel architecture which have shown remarkable computation times for scene representation. Non-convexity, on the other hand, comes from the magnitude-phase parametrization and the neural field architecture. This implies that there are no convergence guarantees, and optimization can end up in poor local minima. The method can also be sensitive to initialization of weights. We highlight, however, that all the neural fields used in our experiments have the same architecture and weights at initialization.

\begin{table}[]
\centering
\begin{tabular}{c|c|c}
\cline{2-3}
 & Cartesian & Radial \\ \hline
Neural Field GPU time (min) & 18 & 43 \\ \hline
LLR CPU time (min) & 8 & 550 \\ \hline
\end{tabular}
\caption{Wall time for neural field and LLR methods on high temporal resolution dataset for Experiment 1 (Cartesian) and Experiment 3.a (radial). Neural field optimization is performed on GPU and the reported time corresponds to the joint resolution for both echoes. LLR's optimization is performed on CPU and the reported time corresponds to the resolution for both echoes.}
\label{tab:total time}
\end{table}

\begin{table}[]
\centering
\begin{tabular}{c|ccc|ccc}
\cline{2-7}
 & \multicolumn{3}{c|}{Cartesian} & \multicolumn{3}{c}{Radial} \\ \hline
Batch size $N_B$ & 1 & 21 & 42 & 1 & 21 & 42 \\ \hline
Memory (GB) & 1.01 & 3.03 & 5.14 & 0.33 & 1.83 & 3.47 \\
Time/epoch (s) & 0.83 & 0.71 & 0.72 & 1.87 & 1.76 & 1.75 \\ \hline
\end{tabular}
\caption{Peak memory consumption during optimization and time per epoch for neural field on Experiments 1 and 3.a at different batch sizes.}
\label{tab:memory}
\end{table}

Although explicit phase unwrapping is not incorporated in the present framework, phase wrapping was not observed in the datasets considered here. More generally, phase wrapping corresponds to a discontinuity that can, in principle, be represented by neural fields, while potential oversmoothing effects are mitigated by the proposed hybrid voxel-based refinement.

The fully-sampled data used in the experiments is collected by gating cardiac phases over many cardiac cycles. A more ambitious step is towards non-gated data and reconstructing the actual spatiotemporal scene. In this case, periodicity is not harnessed in the data. Hence, it must be imposed in some other way, perhaps in the architecture of the neural field. Finally, we note that the proposed approach naturally extends to 4D CPC MRI. In the present 2D experiments, the neural field is evaluated on a grid of size $142\x142\x N_B$, where $N_B$ denotes the temporal batch size. As reported in Table~\ref{tab:memory}, the resulting memory consumption scales approximately linearly with the batch size. A 4D flow reconstruction with batch size one would require evaluating the neural field on a comparably sized spatio-temporal grid, leading to similar memory requirements as those observed at large $N_B$ in the 2D setting. While this implies increased computational cost, the reported results suggest that such evaluations remain feasible on modern GPUs, but with longer runtimes.

\section{Conclusion}

In this paper, we have proposed neural fields for highly accelerated 2D CPC MRI. The neural field uses a magnitude-phase parameterization of the scene and is optimized by solving both velocity encodings in one joint variational problem. In this way, the information of the two echoes interplays to enhance the reconstruction. Additionally, we propose a simple voxel-based postprocessing step that can compensate for the oversmoothing tendency of neural fields arising from implicit regularization and optimization bias. We have validated our method using datasets with different temporal resolutions and with two common sampling strategies, Cartesian and radial.

\section{Experimental section}

In this section, we discuss in detail the mathematical aspects used in this work, from the variational problem with Cartesian and radial data to the neural field and voxel-based parametrized solutions.

\subsection{Variational problem}\label{sec:variational problem}

We consider the following forward model for one echo
\begin{equation}
    f_{c,t} = \bm{K}_{c,t}u_t + \varepsilon_{c,t},
\end{equation}
for coils $c=1,\ldots,N_C$, and times $t=1,\ldots,N_T$. $f_{c,t}$ is the raw k-space data from collected by coil $c$ at time $t$; $\bm{K}_{c,t}$ is the imaging process; $u_t$ is the sought complex-valued image; and $\varepsilon_{c,t}$ is additive Gaussian noise. The imaging process consists of an element-wise multiplication by the sensitivity map $\bm{S}_c$, followed by an FFT $\bm{F}$ and the sampling mask $\bm{M}_t$ for Cartesian data, or a non-uniform FFT $\tilde{\bm{F}}$ for radial data. 

Recall that we have this model for two different echoes $f^0, f^1$. The sensitivity maps $\{\bm{S}_c\}_{c=1}^{N_C}$ are precomputed from these echoes by averaging both in time, and using ESPIRiT with a calibration region of size 16$\x$16 around the center of k-space~\cite{uecker2014espirit}. Therefore, the maps are the same for both echoes. Additionally, at each frame the sampled frequencies are different per echo.

\subsubsection{Cartesian data}

For Cartesian data, the imaging process takes the form 
\[\bm{K}_{c,t} = \bm{M}_t\bm{F}\bm{S}_c, \quad c=1,\ldots,N_C, \quad t=1,\ldots, N_T.\]
The data fidelity term for the variational problem has the form
\begin{equation}
    \mathcal{D}^{\text{cart}}(\bm{K}u, f) = \dfrac{1}{2}\dsum_{t=1}^{N_T}\sum_{c=1}^{N_c}\|\bm{K}_{c,t}u_t - f_{c,t}\|_2^2.
\end{equation}

\subsubsection{Radial data}

For radial sampling, the imaging process is given as follows
\[\bm{K}_{c,t} = \tilde{\bm{F}}_t\bm{S}_c.\]
Here, $\tilde{\bm{F}}_t$ is the non-uniform FFT that samples the points in the k-space trajectory at time $t$. For the loss, we also make use of a density compensation diagonal matrix $d_t$ to account for the oversampled k-space center:
\begin{equation}
\begin{aligned}
    \mathcal{D}^{\text{radial}}(\bm{K}u, f) =& \dfrac{1}{2}\dsum_{t=1}^{N_T}\sum_{c=1}^{N_c}\|d_t(\bm{K}_{c,t}u_t - f_{c,t})\|_2^2\\
    =& \dfrac{1}{2}\dsum_{t=1}^{N_T}\sum_{c=1}^{N_c}\langle(d_t\bm{K}_{c,t})^H(d_t\bm{K}_{c,t})u_t,u_t\rangle\\&-2\text{Re}(\langle u_t, (d_t\bm{K}_{c,t})^Hd_tf_{c,t}\rangle)+c,
\end{aligned}
\end{equation}
with $\cdot^H$ denoting the conjugate transpose. The first term is not computationally efficient to optimize neural fields, since it applies a non-uniform FFT for the forward pass and then its adjoint for the backwards pass. The second term introduces significant benefits in computation time when training the neural field because the Toeplitz kernel $(d_t\bm{K}_{c,t})^H(d_t\bm{K}_{c,t})$ and the adjoint image $(d_t\bm{K}_{c,t})^Hd_tf_{c,t}$ are precomputed once, then, the Toeplitz kernel is applied only once to compute the forward and backwards passes. The non-uniform FFT and density compensation functions are computed with \emph{TorchKbNufft}. 


\subsection{Neural field's architecture}\label{sec:neural field architecture}

The neural field first maps the input $(\bm{x}, t)\mapsto (\gamma_{\bm{x}}(\bm{x}), \gamma_t(t)) \in \R^{2m}$ into a higher dimensional feature vector using two Fourier feature encodings, one for the spatial variable and another for the time variable. These maps are defined as $\gamma_{\bm{x}}(\bm{x}):= (\sin(2\pi \mathbf{B}_{\bm{x}} \bm{x}), \cos(2\pi \mathbf{B}_{\bm{x}} \bm{x}))\in\R^{2m_{\bm{x}}}$ and $\gamma_t(t):= (\sin(2\pi \mathbf{B}_t t), \cos(2\pi \mathbf{B}_t t))\in\R^{2m_t}$, with the sinusoidal functions acting element-wise. The matrices $\mathbf{B}_{\bm{x}}\in\R^{m_{\bm{x}}\times 2}$ and $\mathbf{B}_t\in\R^{m_t\times 1}$ have non-trainable entries sampled from Gaussian distributions $(\mathbf{B}_{\bm{x}})_{ij}\sim \mathcal{N}(0,\sigma_{\bm{x}}^2)$ and $(\mathbf{B}_t)_{ij}\sim \mathcal{N}(0,\sigma_t^2)$. The hyperparameters $\sigma_{\bm{x}}$ and $\sigma_t$  account for the frequencies the neural field can capture; the larger they are, the larger the frequencies can be captured earlier during optimization. For all the experiments, we use $\sigma_{\bm{x}}=0.5$, $\sigma_t=1$, and $m_{\bm{x}}=m_t=32$, leading to a Fourier feature vector $(\gamma_{\bm{x}}(\bm{x}), \gamma_t(t))\in\R^{128}$. This is then the input of a multilayer perceptron with 5 hidden layers with 128 neurons each and $\tanh$ as activation function. Finally, the output layer is obtained by applying a linear transformation leading to an output vector of size 3. The first component of the output is the magnitude, and an exponential activation is applied to that neuron only to ensure its positivity. 




\subsection{Hybrid model: voxel-based postprocessing of neural field}\label{sec:hybrid}

We propose a voxel-based postprocessing of the neural field. This is relevant for phantoms that the neural field is unable to capture, either because of a lack of expressive power or because of converging to a poor local minimum. Due to differentiability and convexity with respect to $u$ of the variational problem in Equation \eqref{eq:variational problem hybrid}, these are updated by imposing the first optimality condition. We get one linear system for each $u$ that is solved with conjugate gradient iterations:
\[ \left( \dsum_{c=1}^{C} (\bm{K}^j_{c,t})^H \bm{K}^j_{c,t} +\lambda_{\text{Hyb}} I \right) u_t^j = \dsum_{c=1}^{C}(\bm{K}^j_{c,t})^Hf_{c,t}^j+ \lambda_{\text{Hyb}} (u^j_{\theta})_t, \quad j=0, 1, \]
for times $t=1,\ldots, N_T$. More specifically, we use a maximum of 30 iterations and a tolerance of $10^{-10}$.

\subsubsection{Ablation study on $\lambda_{\text{Hyb}}$}\label{sec:ablation hyb}

\begin{figure}[htbp]
\centering
\includegraphics[width=0.45\textwidth]{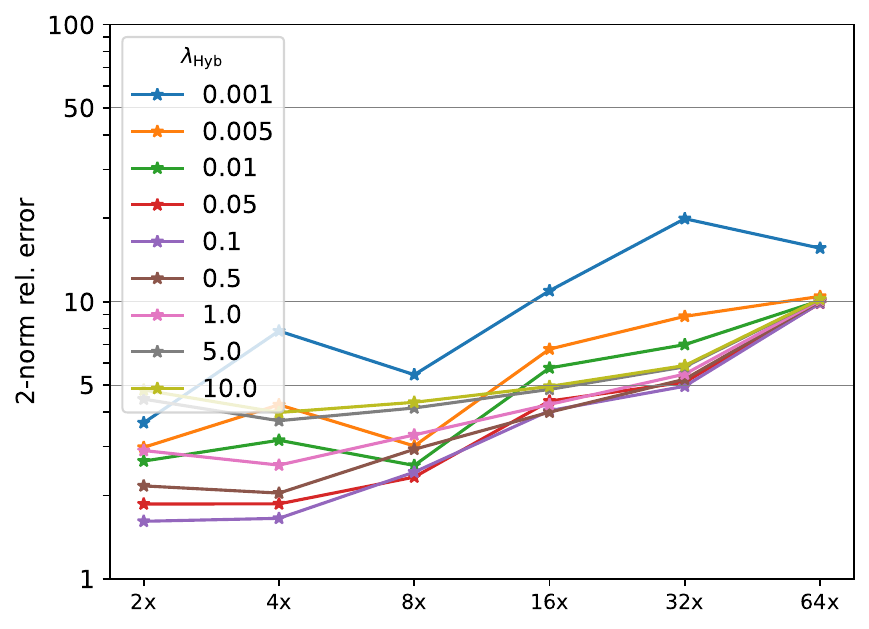}
\caption{Ablation study on $\lambda_{\text{Hyb}}$ for Experiment 1. The 2-norm flow relative error curves are displayed for several values of $\lambda_{\text{Hyb}}$ and acceleration factors. $\lambda_{\text{Hyb}}=10^{-1}$ presents the best overall performance across all factors.}
\label{fig:ablation hyb}
\end{figure}

Here we discuss the choice of the regularization parameter $\lambda_{\text{Hyb}}$ used for the hybrid model. Figure \ref{fig:ablation hyb} displays the results for Experiment 1 with $\lambda_{\text{LLR}}\in\{10^{-3}, 5\times 10^{-3}, 10^{-2}, \ldots, 5, 10\}$ across all acceleration factors, and shows a better overall performance for $\lambda_{\text{Hyb}}=10^{-1}$. Additionally, we observe that for a small value of this parameter, e.g., $\lambda_{\text{Hyb}}=10^{-3}$, the solution gets closer to the unregularized SWS solution yielding large errors, while for large values of this parameter, e.g., $\lambda_{\text{Hyb}}=10$, the solution barely moves from the neural field's prediction, as expected.

\subsection{Sensitivity weighted solution (SWS)}\label{sec:sws}

This solution is obtained by solving two independent variational problems with no regularization, one for each encoding $f^0, f^1$:
\begin{equation}
u_{\text{SWS}}^j:=\arg\min_{u\in\mathbb{C}^{N\x N_T}}\mathcal{D}(\bm{K}^ju,f^j), \quad j=0,1.
\end{equation}

The first-order optimality condition for the Cartesian loss leads to the following linear system whose solution is the SWS:
\begin{equation}
\dsum_{c=1}^{C} (\bm{K}^j_{c,t})^H \bm{K}^j_{c,t} (u_{\text{SWS}}^j)_t = \dsum_{c=1}^{C}(\bm{K}^j_{c,t})^Hf^j_{c,t},
\end{equation}
for times $t=1,\ldots, N_T$. This solution is analogous to the zero-filled solution, but for parallel imaging, where sensitivity coils need to be accounted for. Numerically, this system is solved with conjugate gradient iterations. 

\subsection{Locally low-rank regularized solution}\label{sec:llr}
We solve one variational problem for each flow encoding $f^0, f^1$:
\begin{equation}
u_{\text{LLR}}^j:=\arg\min_{u\in\mathbb{C}^{N\x N_T}}\mathcal{D}(\bm{K}^ju, f^j) + \lambda_{\text{LLR}}\dsum_{i=1}^P\|\bm{P}_iu\|_*, \quad j=0,1.
\end{equation}
The second term is the locally low-rank regularizer, where $\bm{P}_i$ extracts a small patch of $u$ of size $8\times8\times N_T$ and reshapes it into the Casorati matrix, $\|\cdot\|_*$ is the nuclear norm acting on non-overlapping patches by penalizing their rank, and $\lambda_{\text{LLR}}\geq 0$ is the regularization parameter. Numerically, the problem is solved using FISTA over 30 iterations, as implemented in the BART toolbox for computational MRI~\cite{uecker2015berkeley}.

\subsubsection{Ablation study on $\lambda_{\text{LLR}}$}\label{sec:ablation llr}

\begin{figure}[htbp]
\centering
\includegraphics[width=0.45\textwidth]{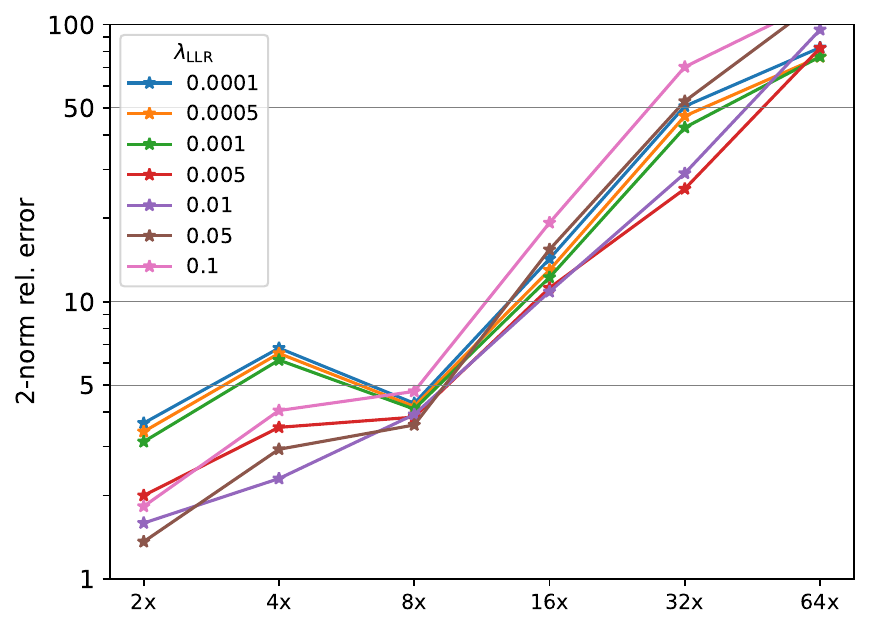}
\caption{Ablation study on $\lambda_{\text{LLR}}$ for Experiment 1. The 2-norm flow relative error curves are displayed for several values of $\lambda_{\text{LLR}}$ and acceleration factors. $\lambda_{\text{LLR}}=10^{-2}$ presents the best overall performance across all factors.}
\label{fig:ablation llr}
\end{figure}

Here we discuss the choice of the regularization parameter $\lambda_{\text{LLR}}$ used for the locally low-rank model. Figure \ref{fig:ablation llr} displays the results for Experiment 1 with $\lambda_{\text{LLR}}\in\{10^{-4}, 5\times 10^{-4}, 10^{-3}, \ldots, 5\times 10^{-2}, 10^{-1}\}$ across all acceleration factors, and shows a better overall performance for $\lambda_{\text{LLR}}=10^{-2}$. More importantly, the figure shows that LLR works well for acceleration factors up until 8$\x$, however, the method fails to capture temporal coherence for large factors, leading to large errors in the predicted flow: with such limited k-space coverage per frame, the patch-based low-rank assumption becomes unreliable, and LLR struggles to enforce global temporal coherence.

\subsection{Metrics}\label{sec:metrics}

The flow at time $t$ is computed as
\[Q_t:= |A_t|\dfrac{1}{N}\dsum_{\bm{x}\in A_t}v(\bm{x}, t), \quad t=1,\ldots, N_T,\]
where $A_t$ is the manually segmented aorta from the reference image at time $t$, $v$ is the phase difference, and $\bm{x}$ are voxels in the aorta. To measure errors we compute the 2-norm, $\infty$-norm, and the overall flow relative errors:
\[\dfrac{\|\bm{Q}-\bm{Q}^*\|_2}{\|\bm{Q}^*\|_2}\times 100 \%, \quad \dfrac{\|\bm{Q}-\bm{Q}^*\|_{\infty}}{\|\bm{Q}^*\|_{\infty}}\times 100 \%,\] 
\[ \dfrac{| \sum_t Q_t - \sum_t Q_t^*|}{|\sum_t Q_t^*|}\times 100 \%,\]
where $\bm{Q}=[Q_1,\ldots, Q_{N_T}]^T$ is the predicted solution and $\bm{Q}^*$ the ground truth flow.

\subsection{Embedding problem for neural field}\label{sec:embedding}

\begin{figure}[htbp]
    \centering
    \includegraphics[width=0.45\linewidth]{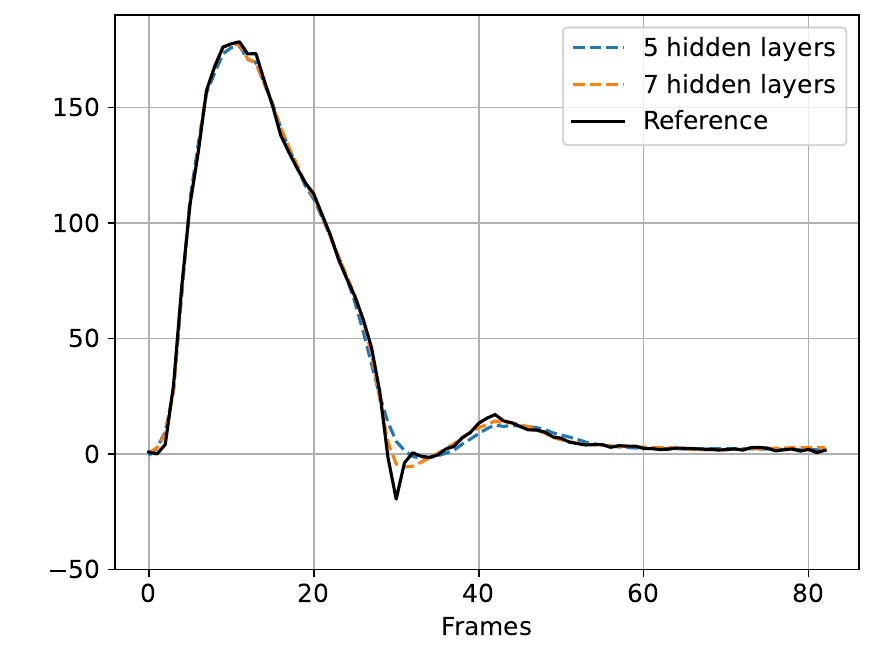}
    \caption{Embedding problem for high temporal resolution data using 5 and 7 hidden layers, and Fourier encoding with $\sigma_{\bm{x}}=0.5$, $\sigma_t=1$. The studied architectures do not capture the negative peak at frame 30 even when directly fitting the reference image.}
    \label{fig:embedding}
\end{figure}

In this section, we perform a reference experiment in which the neural field directly fits the high temporal resolution reference images $u^0, u^1$. The goal is to address the representation power of neural fields under different architectures. Specifically, we solve
\[\min_{\theta}\dfrac{1}{2}\left(\|u_{\theta}^0-u^0\|_2^2+\|u_{\theta}^1-u^1\|_2^2\right).\]
We try the network with Fourier encoding and $\tanh$ using 5 and 7 hidden layers and varying $\sigma_t=1, 10$ with fixed $\sigma_{\bm{x}}=0.5$. We also try a SIREN architecture with sinusoidal activation function \cite{sitzmann2020implicit}, 6 and 8 hidden layers, and with frequency parameters $\omega_{\bm{x}}=30$ and $\omega_t=30, 60$. In SIREN, these parameters are used in the first layer by mapping $(\bm{x},t)\mapsto \sin(\omega_{\bm{x}} W_1\bm{x} + \omega_tW_2 t)$. The extra layer in SIREN is employed to mimic the Fourier encoding layer. In these architectures, the hyperparameters $\sigma_t$ and $\omega_t$ control the temporal frequency content of the neural field, with larger values enabling higher-frequency representations but also increasing sensitivity to noise in ill-posed inverse problems. These are varied to see if larger frequencies in time can capture the sharp feature shown in Figure \ref{fig:high res flow}. Results are summarized in Table \ref{tab:embedding}, where the PSNR of magnitude and 2-norm relative error of flow are shown. It can be seen that SIREN performs worse despite having an additional layer of trainable parameters. The predicted flow for Fourier encoding + $\tanh$ with $\sigma_{\bm{x}}=0.5, \sigma_t=1,$ and 5 and 7 layers is shown in Figure \ref{fig:embedding}, where a similar smoothing as in Figure \ref{fig:high res flow} is observed around frame 30 even in this simple case of direct fitting. This indicates that the neural field does not fully capture certain details of the reference image, likely due to a combination of spectral bias and the non-convex optimization landscape. Consequently, a similar effect is expected in the inverse problem, where the voxel-based component of the hybrid model can compensate for these shortcomings. This observation highlights the role of the hybrid model. We emphasize that this experiment does not imply that the neural field cannot represent the reference image exactly: such a solution may exist, but the training becomes trapped in suboptimal local minima. 

\begin{figure}[htbp]
    \centering
    \includegraphics[width=0.45\linewidth]{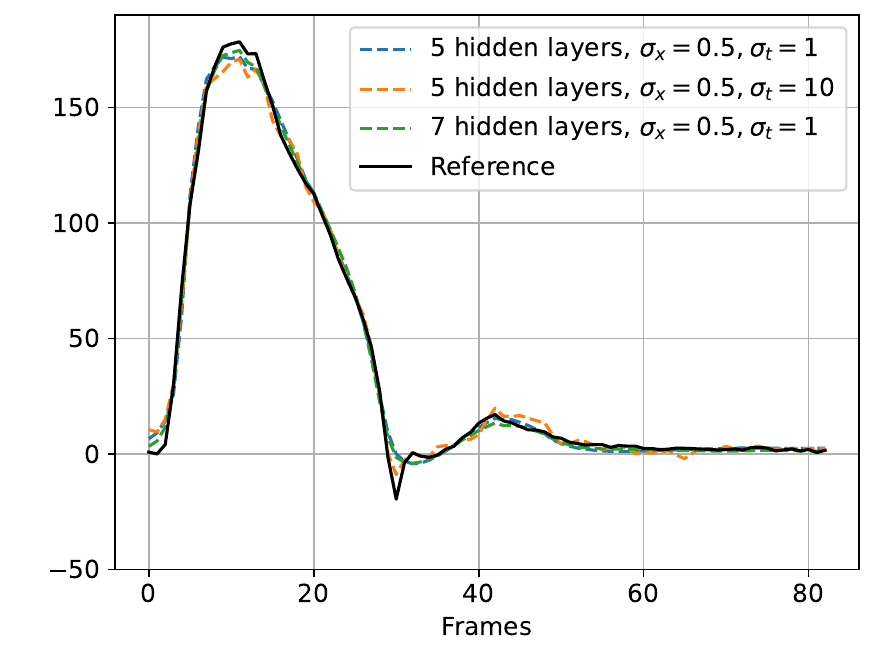}
    \caption{Different architectures for Experiment 1 at an acceleration factor of 32$\x$. We use 5 and 7 hidden layers, and Fourier encoding with $\sigma_t=1, 10$. More hidden layers can lead to better results at the cost of higher time and memory consumption.}
    \label{fig:embedding experiment 1}
\end{figure}

We finish this section by mentioning what happens in Experiment 1 when we increase $\sigma_t$ from 1 to 10 and the number of hidden layers from 5 to 7. The flow curves obtained at an acceleration factor of 32$\x$ are shown in Figure \ref{fig:embedding experiment 1}. While increasing the bandwidth to $\sigma_t=10$ improves the embedding relative error (from 4.9\% to 3.9\%), we observe that in the inverse problem setting, higher bandwidths introduce increased oscillatory behavior in the reconstructed flow curves, particularly at high undersampling factors. For example, at an acceleration factor of 32$\x$ the improvement in relative flow error is marginal from $6.0\%$ with $\sigma_t=1$ to $5.7\%$ with $\sigma_t=10$, while at an acceleration factor of 64$\x$ the higher-bandwidth encoding leads to noisier flow estimates and increased error, from $9.9\%$ to $12.9\%$. Increasing the number of layers to 7 while keeping $\sigma_t=1$, at an acceleration factor of 32$\x$ reduced the relative flow error from 6.0\% to 4.6\%. This improvement comes at a moderate increase in runtime (from 18 to 22 minutes) and memory usage (from 5.14 GB to 5.45 GB). Although further hyperparameter tuning may yield marginal improvements, we adopt a fixed configuration to focus on robustness and reproducibility rather than exhaustive optimization.

\begin{table}[htbp]
\centering
\begin{tabular}{|c|c|c|c|c|}
\hline
Method & Hyperparameter & Hidden layers & Rel. error & PSNR \\ \hline
\multirow{4}{*}{\begin{tabular}[c]{@{}c@{}}Fourer \\ encoding \\ + $\tanh$\end{tabular}} & $\sigma_{\bm{x}}=0.5, \sigma_t=1$ & 5 & 4.9\% & 38.69 \\ \cline{2-5} 
 & $\sigma_{\bm{x}}=0.5, \sigma_t=1$ & 7 & $\mathbf{3.3\%}$ & 39.68 \\ \cline{2-5} 
 & $\sigma_{\bm{x}}=0.5, \sigma_t=10$ & 5 & 3.9\% & 38.65 \\ \cline{2-5} 
 & $\sigma_{\bm{x}}=0.5, \sigma_t=10$ & 7 & 3.5\% & $\mathbf{39.81}$ \\ \hline
\multirow{4}{*}{SIREN} & $\omega_{\bm{x}}=30, \omega_t=30$ & 6 & 6\% & 36.53 \\ \cline{2-5} 
 & $\omega_{\bm{x}}=30, \omega_t=30$ & 8 & 5.8\% & 37.90 \\ \cline{2-5} 
 & $\omega_{\bm{x}}=30, \omega_t=60$ & 6 & 5.4\% & 34.24 \\ \cline{2-5} 
 & $\omega_{\bm{x}}=30, \omega_t=60$ & 8 & 5.7\% & 36.89 \\ \hline
\end{tabular}
\caption{Ablation study on the neural field's architecture for the embedding problem. We vary the time-frequency hyperparameters $\sigma_t$ (for Fourier encoding) and $\omega_t$ (for SIREN) and the number of hidden layers. 2-norm relative error on the flow and PSNR on the magnitude are reported. The largest PSNR and smallest flow error are in bold.}
\label{tab:embedding}
\end{table}

\section*{Acknowledgments}
PA is supported by a scholarship from the EPSRC Centre for Doctoral Training in Statistical Applied Mathematics at Bath (SAMBa), under the project EP/S022945/1. MJE acknowledges support from the EPSRC (EP/T026693/1, EP/V026259/1, EP/Y037286/1). MM acknowledges support from Cancer Research UK Cambridge Centre [CTRQQR-2021/100012] and Cambridge Experimental Cancer Medicine Centre (ECMC) [ECMCQQR-2022/100003]. CBS acknowledges support from the Royal Society Wolfson Fellowship, the EPSRC advanced career fellowship EP/V029428/1, the EPSRC programme grant EP/V026259/1, the Wellcome Innovator Awards 215733/Z/19/Z and 221633/Z/20/Z, the EPSRC funded ProbAI hub EP/Y028783/1. MJE and CBS also acknowledge support from the European Union Horizon 2020 research and innovation programme under the Marie Skłodowska-Curie grant agreement REMODEL. This research was also supported by the NIHR Cambridge Biomedical Research Centre (NIHR203312). The views expressed are those of the author(s) and not necessarily those of the NIHR or the Department of Health and Social Care.
The code and data used in this study will be made publicly available upon publication.


\newpage
\bibliographystyle{unsrt}  
\bibliography{wileyNJD-Chicago}

\end{document}